\documentclass[twocolumn,secnumarabic,amssymb, nobibnotes,  superscriptaddress,aps, prl]{revtex4-2}

\usepackage{graphicx}
\usepackage{amssymb}
\usepackage{amsmath}
\usepackage{color}
\usepackage{graphicx}
\usepackage{dcolumn}
\usepackage{bm}
\usepackage{natbib}
\usepackage[colorlinks=true,citecolor=blue]{hyperref}

\begin{document}

\preprint{APS/123-QED}

\title{Energetics of microwaves probed by double quantum dot absorption}

\author{Subhomoy Haldar}
\email{subhomoy.haldar@ftf.lth.se}
 \address{NanoLund and Solid State Physics, Lund University, Box 118, 22100 Lund, Sweden}
 \author{Harald Havir}
 \address{NanoLund and Solid State Physics, Lund University, Box 118, 22100 Lund, Sweden}
 \author{Waqar Khan}
 \address{NanoLund and Solid State Physics, Lund University, Box 118, 22100 Lund, Sweden}
 \address{Currently at VTT Technical Research Centre of Finland, Box 1000, 02044, Finland}
 \author{Sebastian Lehmann}
 \address{NanoLund and Solid State Physics, Lund University, Box 118, 22100 Lund, Sweden}
\author{Claes Thelander}
 \address{NanoLund and Solid State Physics, Lund University, Box 118, 22100 Lund, Sweden}
\author{Kimberly A. Dick}
 \address{NanoLund and Solid State Physics, Lund University, Box 118, 22100 Lund, Sweden}
 \address{Center for Analysis and Synthesis, Lund University, Box 124, 22100 Lund, Sweden}
\author{Ville F. Maisi}
 \email{ville.maisi@ftf.lth.se}
 \address{NanoLund and Solid State Physics, Lund University, Box 118, 22100 Lund, Sweden}
\date{\today}
\begin{abstract}
We explore the energetics of microwaves interacting with a double quantum dot photodiode and show wave-particle aspects in photon-assisted tunneling. The experiments show that the single-photon energy sets the relevant absorption energy in a weak-drive limit, which contrasts the strong-drive limit where the wave amplitude determines the relevant-energy scale and opens up microwave-induced bias triangles. The threshold condition between these two regimes is set by the fine-structure constant of the system. The energetics are determined here with the detuning conditions of the double dot system and stopping-potential measurements that constitute a microwave version of the photoelectric effect.
\end{abstract}
\maketitle

Microwaves have an important role in solid-state quantum systems in mediating interactions, control, and performing readout in quantum technology~\cite{Gu2017}. One particularly interesting subset of the field is semiconductor-superconductor hybrids where the microwave signals interact with low-dimensional semiconductors enabling coherent interactions together with electron transport~\cite{Burkard2020}. One of the central nanostructures here is a double quantum dot (DQD)~\cite{childress2004, vanderWiel2002,chatterjee2021}. DQDs have been used for photon emission~\cite{liu2014, stockklauser2015} up to masing regime ~\cite{peiqing2011,liu2015,Gullans2015, agarwalla2019}, coherent interactions~\cite{frey2012, mi2017, stockklauser2017, samkharadze2018} including ultra-strong coupling limit~\cite{scarlino2021}, and microwave spectroscopy~\cite{stafford1996, oosterkamp1998, vanZanten2020}. The recent theory by Wong and Vavilov~\cite{wong2017}, predicted, and the following experiment by Khan \textit{et al.}~\cite{khan2021} demonstrated that a cavity-coupled DQD yields an efficient photodiode in the microwave domain. Detection of microwave photons using DQDs is encouraging because of the capabilities offered due to the continuous mode of device operation with voltage-controlled energy levels~\cite{ghirri2020,gustavsson2007,zenelaj2022}. However, most of these previous reports describing DQDs as microwave absorbers lack an explicit discussion of wave-particle interplay during the absorption process. It is often considered that the relevant threshold for the single-photon picture is when the resonator is loaded on average by one photon or less.

\begin{figure}[b]
\includegraphics[width=3.1in]{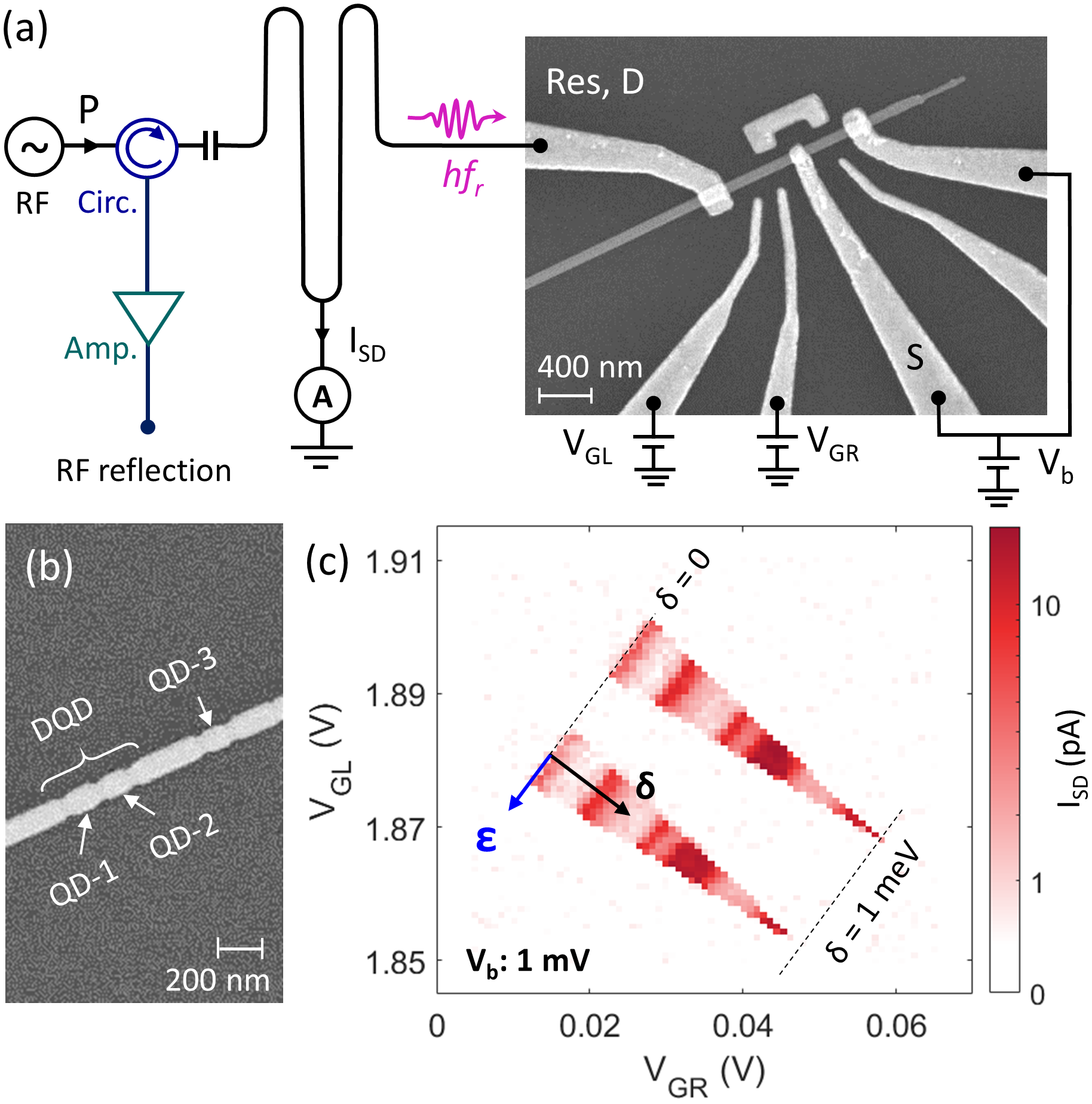}
\caption{\label{fig1}(a): Scanning electron microscope image of the nanowire device and schematic diagram of the coplanar resonator and measurement setup. The central conducting line of the resonator connects the drain lead D of the DQD and allows to measure $I_\mathrm{SD}$ via a voltage node point of the resonance mode. (b): The nanowire used for the device fabrication before removing the GaSb shell shows the individual dots. (c): Bias voltage $V_b$ driven current $I_\mathrm{SD}$ as a function of $V_\mathrm{GL}$ and $V_\mathrm{GR}$ in absence of a microwave-drive.}
\end{figure}

In this letter, we however show that the single-photon absorption picture~\cite{childress2004,vanderWiel2002,khan2021} sustains up to the cavity photon number $n_r^c=9/16\alpha$ = 490, set by the effective fine-structure constant $\alpha = \frac{1}{4}Z_0G_0$ with characteristic impedance $Z_0$ of the resonator and conductance quantum $G_0$. We probe the energetics~\cite{bergenfeldt2014,viennot2014} of the microwaves interacting with a DQD photodiode and determine the detuning conditions leading to photon-activated transport. These results are complemented with stopping-potential measurements that constitute a microwave version of the photoelectric effect showing the photo-excited electrons obtain nearly the full energy of the photons in low-drive limit.

\begin{figure*}
\includegraphics[width=6.9in]{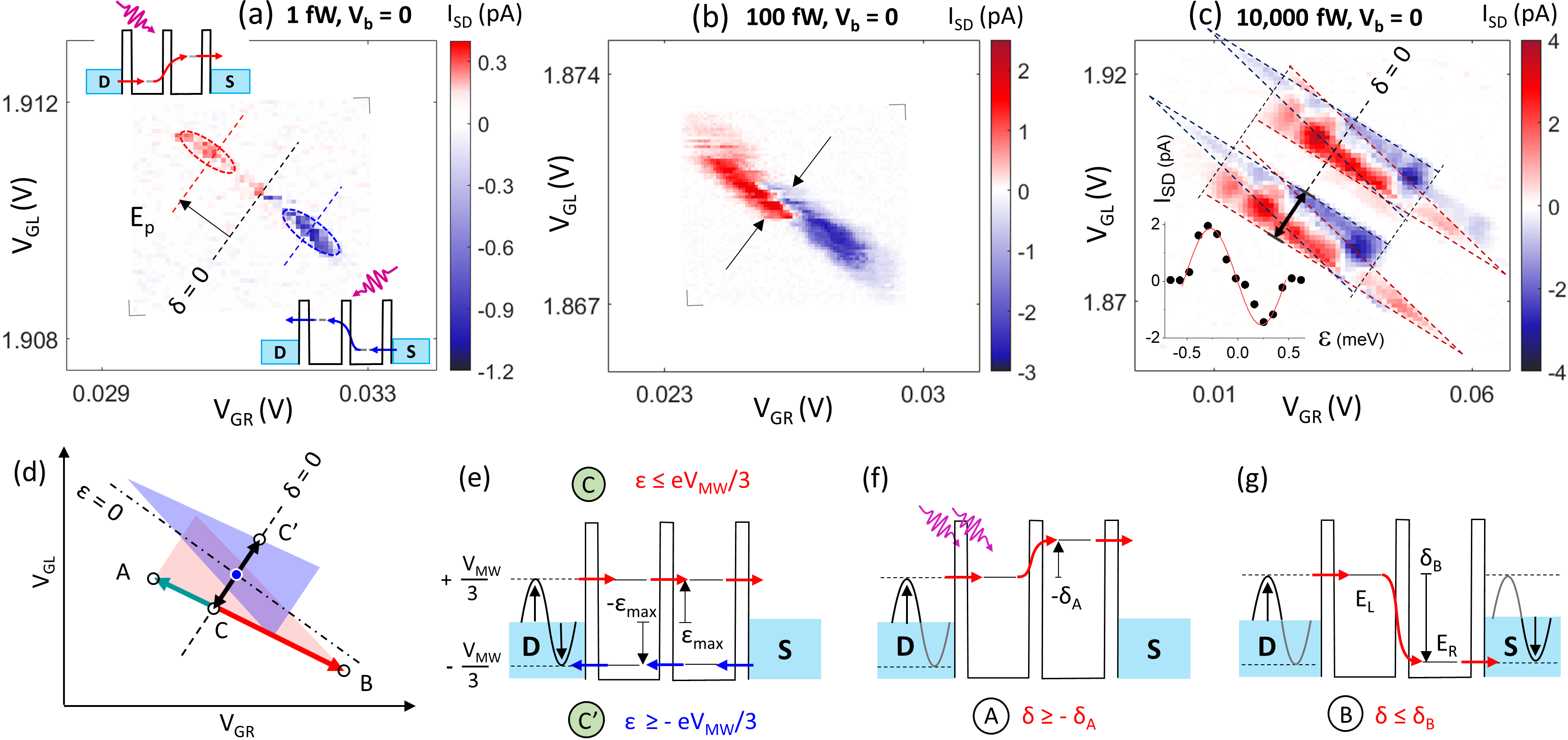}
\caption{\label{fig2} (a)-(c): Photocurrent as a function of $V_{\mathrm{GR}}$ and $V_{\mathrm{GL}}$ with $V_b$ = 0 and $P=$ 1~fW (a), 100 fW (b), and 10~000 fW (c). Inset of figure (c) shows $I_{\mathrm{SD}}$ as a function of $\varepsilon$ at $\delta=0$. Dashed lines drawn in panels a and c are guides for the eye. (d): Schematic diagram of the overlapping microwave-induced-triangles with the critical points A, B, C, and C' for the analysis. The triangles extend both in the positive and negative detuning $\delta$ directions along with a finite spread in $\varepsilon$ axis (line CC' for the $\delta = 0$ case). (e)-(g): Energy band diagrams and threshold conditions for points C and C' (e), point A with negative detuning $\delta = -\delta_A$ (f), and point B with positive detuning $\delta = \delta_B$ (g). The sinusoidal curves of (e-g) indicate the amplitude ($V_\mathrm{MW}/3$) which is capacitively coupled across the junctions. Thus, the energy levels of the two dots and the Fermi energy of the drain are coupled to this voltage amplitude, where the Fermi energy of the source sets the reference potential.}
\end{figure*}

The hybrid device used in the experiments is realized by coupling a DQD system to a coplanar waveguide resonator, Fig.~\ref{fig1}a. Here, a crystal phase defined DQD is formed by two zincblende islands separated by three wurtzite barriers in an InAs nanowire ~\cite{Luna2015,barker2019,khan2021}. A GaSb shell, selectively grown on zincblende segments and removed before making contacts, helps to locate the dots as seen in Fig.~\ref{fig1}b. Plunger gate voltages ($V_{\mathrm{GL}}$, $V_{\mathrm{GR}}$) control the electrochemical energies ($E_L$, $E_R$) of the respective dots. The drain contact couples directly to a coplanar resonator made of 100 nm thick Nb film. The resonator has a fundamental resonance at $f_r$ = 6.715 GHz corresponding to photon energy $E_P$ = 27.7 $\mu$eV, and a loaded quality factor of $Q_L$ = 1320 with lumped-element capacitance $C_r\approx$ 540 fF and inductance $L_r\approx$ 1000 pH. Details of our RF circuit and power calibration may be found in Ref.~\onlinecite{khan2021}. For standard transport, we apply a bias voltage $V_b$ to the source contact and measure current $I_\mathrm{SD}$ from the voltage node point in the middle of the resonator. The nanowire also contains an additional dot (QD-3) for charge sensing, which is not used in our present study. To avoid contribution from QD-3, we apply the same bias to the other side of QD-3, Fig.~\ref{fig1}a. Measurements are performed in a dilution fridge with a base temperature of 10 mK.

\par We first characterize the properties of the DQD by measuring a charge stability diagram without a microwave drive (Fig.~\ref{fig1}c). With $V_b$ = 1 mV applied on the DQD, sequential tunneling of electrons results in finite bias triangles (FBTs) at the charge triple points~\cite{vanderWiel2002}. To specify the energy level positions, we define two local coordinates $\delta$ = ($E_L-E_R$) and $\varepsilon$ =($E_L+E_R$)/2 representing difference and average energy of electronic states in the two dots (Fig.~\ref{fig1}c). Both $\delta$ and $\varepsilon$ change linearly with the applied gate voltage, $\delta = \alpha_\delta^R\times\Delta V_{\mathrm{GR}}$ and $\varepsilon = \alpha_\varepsilon^R\times\Delta V_{\mathrm{GR}}$. This yields us the lever arms, $\alpha_{\delta}^R$ = 29.5 $\mu$eV/mV and $\alpha_{\varepsilon}^R$ = 75.3 $\mu$eV/mV along the two directions indicated in Fig.~\ref{fig1}c with black and blue arrows respectively~\cite{khan2021, vanderWiel2002}. From the charge stability diagram, we estimate the charging energies $E_\mathrm{C,L}$ = 1.9 meV, $E_\mathrm{C,R}$ = 2.2 meV~\cite{barker2019}, and restrict our analysis to $\delta, \varepsilon  < E_\mathrm{C}$ to allow only one excess electron in the DQD.

\par We now turn into the microwave response of the circuit. A continuous microwave tone is sent into the resonator via a capacitively coupled input-port shown in Fig.~\ref{fig1}a. The DQD resides at the voltage anti-node on the other end of the $\lambda/2$ resonator. The electrical dipole moments of the DQD charge states hence interact with voltage oscillations at the end of the resonator~\cite{frey2012}. Figures~\ref{fig2}a - c present the measured $I_\mathrm{SD}$ through the DQD at power $P = 1, 100$ and 10\:000~fW respectively. Panel a with the lowest power shows finite $I_\mathrm{SD}$ at two points with polarity reversal. These points correspond to the cases where the energy difference of the bonding and anti-bonding states of the DQD, given by $E = \sqrt{\delta^2+(2t)^2}$, matches the single-photon energy $E_P$, which corresponds to the 'microwave as a particle' scenario~\cite{khan2021}. The energy diagrams in the insets illustrate this with vanishing hybridization. With larger microwave power, panels b and c, the two points increase in size and eventually form microwave induced triangles (MWTs) with the same shape as the FBTs of Fig.~\ref{fig1}c. Interestingly, in contrast to the low-power limit, the active regions extend now both in the positive and negative $\delta$ directions. The formation of these MWTs in electronic transport was not presented in the earlier reports even under high microwave
power~\cite{vanderWiel2002, oosterkamp1998, wacker1997, tien1963}. The faint photo-current signal outside the MWTs of Fig.~\ref{fig2}c arises most likely due to the co-tunneling of charges across the transparent QD-lead barrier of the DQD~\cite{vanderWiel2002}.

\par To understand the energy scales governing the formation of the MWTs, we repeat in Fig.~\ref{fig2}d a microwaves-driven stability diagram for a high-power case schematically together with the key detuning points for our analysis marked with A, B, C and C'. We begin by identifying the $(\delta, \varepsilon) = (0,0)$ point marked in blue based on the symmetries of the two overlapping MWTs. Along the $\delta = 0$ line, we obtain a finite $I_\mathrm{SD}$ up to the points where the energy $\varepsilon_\mathrm{max}$ of adding an electron into the DQD, point C, or the energy of removing an electron from the dot $-\varepsilon_\mathrm{max}$, point C', exceeds what is available from the microwave-drive. Figure~\ref{fig2}e presents the corresponding energy band diagrams. The data of Fig.~\ref{fig2}c yields the value $\varepsilon_\mathrm{max} = 520\ \mathrm{\mu eV}$ for $P = 10$ pW. In a steady state, photocurrent depends on the fraction of time when the DQD energy levels ($E_L$, $E_R$) remain below the Fermi energy of the drain. Thus, the approximately sinusoidal behavior of $I_\mathrm{SD}$, as seen in the inset of Fig.~\ref{fig2}c, represents the microwave waveform interacting with the Fermi energy of the drain.

\begin{figure}[t]
\includegraphics[width=3.3in]{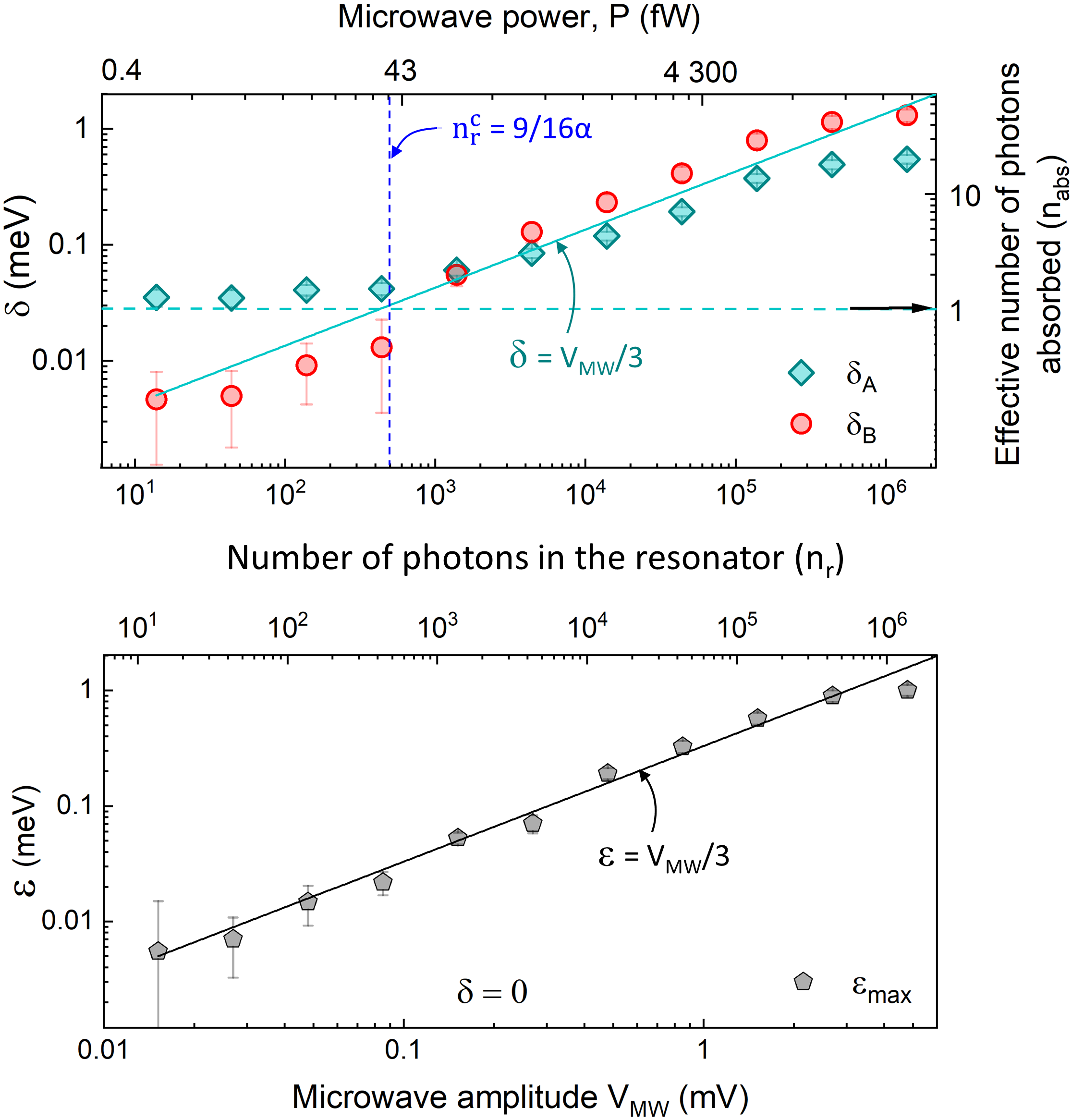}
\caption{\label{fig3}(a): Detuning $\delta_A$ and $\delta_B$ as a function of microwave power $P$. Here, the solid green line corresponds to $\delta = V_\mathrm{MW}/3$ and the dashed line $\delta = E_P$. The corresponding number of photons ($n_\mathrm{abs}$ = $\delta_A/E_P$) contributing to an absorption event is also shown on the right-hand side. (b): Energy $\varepsilon_\mathrm{max}$ of the MWTs as a function of $P$. The microwave amplitude $V_\mathrm{MW}$ corresponding to the input powers are also indicated.}
\end{figure}
\begin{figure*}
\includegraphics[width=6.95 in]{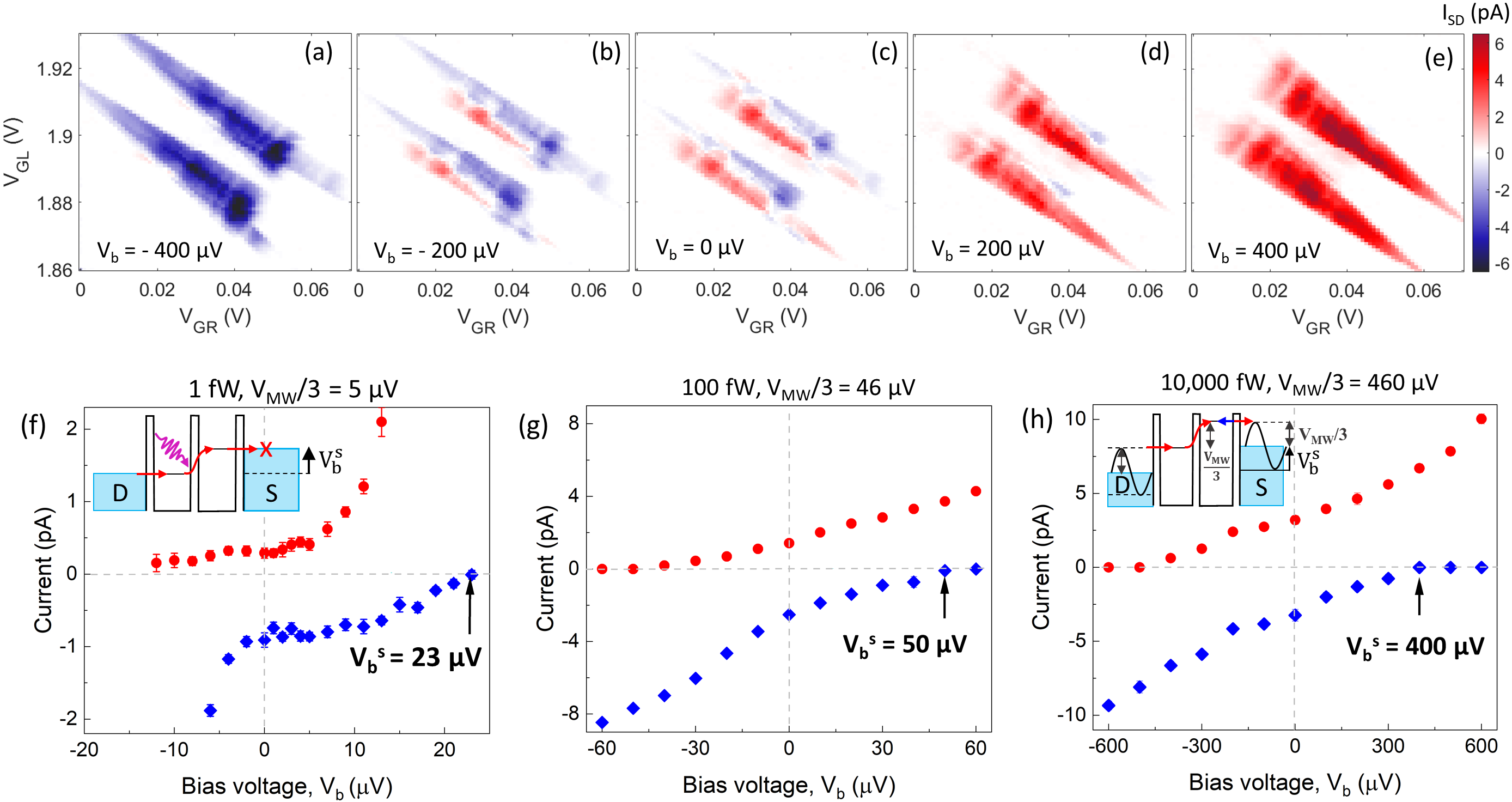}
\caption{\label{fig4} (a)-(e): Photocurrent $I_\mathrm{SD}$ as a function of the plunger gate voltages for $P = 10$ pW and applied bias voltage $V_b$ ranging from -400 to 400 $\mu V$.
(f)-(h): Positive (red) and negative (blue) $I_\mathrm{SD}$ as a function of $V_b$ for $P = $ 1, 100, 10~000 fW respectively.}
\end{figure*}

\par Moving from point C of Fig.~\ref{fig2}d towards either A or B along the edge of the red MWT keeps the energy level $E_L$ fixed while changing $E_R$. Moving from C towards A introduces an energy cost $-\delta$ needed for the interdot transition from left QD to the right one reaching its maximum $-\delta = \delta_A$ at point A as presented in Fig.~\ref{fig2}f. From Fig.~\ref{fig2}c, we determine $\delta_A = 370\ \mathrm{\mu eV}$ as the largest amount of energy provided by the microwave drive with $P = 10$ pW. 

When moving from point C instead to point B, energy is gained in the interdot transition. Here, in addition to photons, phonons play a major role in absorbing the energy~\cite{liu2014,stockklauser2015}. Interestingly the maximum energy $\delta = \delta_B = 880\ \mathrm{\mu eV}$ at point B as shown in panels c and g is larger than $\varepsilon_\mathrm{max}$. This implies that the electron enters the S reservoir at an energy $\delta_B - \varepsilon_\mathrm{max} = 360\ \mathrm{\mu eV}$ below the Fermi energy, see Fig.~\ref{fig2}g. This is possible only if part of the microwave drive appears across the right QD - S barrier. A simple estimate that the RF-amplitude $V_\mathrm{MW}$ is divided roughly equally over the three junction-capacitances is supported by the finding that the corresponding energy costs, $\varepsilon_\mathrm{max}$, $\delta_A$ and $(\delta_B - \varepsilon_\mathrm{max})$ are approximately equal.

\par Figure~\ref{fig3} presents $\delta_A$, $\delta_B$ and $\varepsilon_\mathrm{max}$ as a function of $P$. As seen in Fig.~\ref{fig3}, both $\delta_A$ and $\delta_B$ as well as $\varepsilon_\mathrm{max}$  all increase proportional to $P^{1/2}$ over two orders of magnitude in the power. This indicates the amplitude $V_\mathrm{MW} \propto P^{1/2}$ sets the relevant energy for the transport. To estimate $V_\mathrm{MW}$ across the DQD, we use the result of Ref.~\citealp{sage2011}, which considers the standing wave in the resonator by circulating the input power for $Q_L$ oscillation periods and hence increasing the amplitude to
\begin{equation}\label{eq1}
    V_\mathrm{MW}=\left(\frac{4Q_L^2 Z_0 }{Q_{\mathrm{ex}}}\times P\right)^{1/2}
\end{equation}
Here, $Z_0 \approx 60\, \Omega$ is the resonator impedance, and $Q_{\mathrm{ex}}$ = 1820 is the external quality factor. As shown by the solid lines of Fig.~\ref{fig3}, one-third of the estimated microwave-amplitude per junction, $eV_\mathrm{MW}/3$, catches well the relevant energy costs in the large $P$ region. Correspondingly the number of photons stored in the resonator is obtained as the ratio of energy stored in the resonator to the photon energy, $n_r = \frac{1}{2}C_rV_\mathrm{MW}^2/E_P$, which we depict in Fig.~\ref{fig3}.

\par For $P < 100$ fW, the inter-dot absorption energy $\delta_A$ saturates to $E_P$ (Fig.~\ref{fig3}a). This is the smallest energy quanta that can be absorbed from the microwave drive. The saturation takes place at $eV_\mathrm{MW}/3 \lesssim E_P$ when the energy related to the wave amplitude is not sufficient to provide energy greater than the photon energy. For $\delta_B$ and $\varepsilon_\mathrm{max}$, the saturation is not present. These two energies are related to the tunneling between one of the discrete QD energy levels and the continuum of states in either source or drain in contrast to tunneling between the discrete energy levels in the interdot process. 

The threshold condition, $eV_\mathrm{MW}/3 = E_P$, yields the corresponding cavity photon number as $n_r^c = \frac{1}{2}C_r(3 E_P/e)^2/E_P = 9 \alpha /16$, where $\alpha = \frac{1}{4}Z_0 G_0$ is the fine-structure constant of the system. The blue dashed line of Fig.~\ref{fig3}a marks the threshold value $n_r^c$ = 490, which is in good agreement with the threshold in our experiment. Thus, the single-photon absorption picture holds up to the large photon number of $n_r^c$ = 490 set by the resonator impedance level via $\alpha$ together with the prefactor of $(m/4)^2$, where $m$ = 3 arises from the number of junctions over which the resonator voltage is divided.

\par An alternative approach to determine the absorption energetics is to apply a finite bias $V_b$ to stop the photo-generated current. This is analogous to the famous photoelectric effect of optical photons. In Figs.~\ref{fig4}a-e, we show the effect of $V_b$ on the photoresponse. The middle panel repeats the results at no applied bias. Once we apply a positive $V_b$, the negative MWT starts to shrink and finally vanishes as seen in Figs.~\ref{fig4}d and e. Similarly, the positive MWT diminishes for $V_b<0$ as presented in ~\ref{fig4}a and b. Figures~\ref{fig4}f-h show the maximum values of positive and negative current as a function of $V_b$ for three microwave powers. The stopping-potential $V^\mathrm{s}_b$ marked with black arrows is the point where the reverse direction current $I_\mathrm{SD}<0$ vanishes. 

At the low-drive limit of Fig.~\ref{fig4}f, we have $eV_b^\mathrm{s} \approx E_P$. The amplitude $V_\mathrm{MW}$ is low and hence the energy gain of the transported electrons is set by the single-photon energy $E_P$ of the interdot process as depicted in the inset. For the higher power of Figs.~\ref{fig4}g and h, stopping-potential increases and follows $V_b^\mathrm{s} \propto P^{1/2}$, which again reflects that the amplitude rather than the photon energy sets the relevant energy-scale. For $P=100$ and $10000$ fW, we have $V_b^s=V_\mathrm{MW}/3$, i.e., the energy gained out of only one of the three tunneling processes. We interpret this to arise from the opposite direction current dominating the transport before reaching $V_b^\mathrm{s} = eV_\mathrm{MW}$ where energy would be gained in all of the three tunneling events. As depicted in the inset of Fig.~\ref{fig4}h, the threshold for the competing opposite current is at $V_b^s=V_\mathrm{MW}/3$ when the amplitude across the right quantum dot and source overcomes the energy cost for tunneling into the DQD as indicated with the blue arrow.

In conclusion, we studied the energetics of a microwave signal absorbed by a DQD-photodiode and explored the wave-particle interplay during the absorption process. At the threshold condition, the number of photons stored in the resonator is solely determined by the fine-structure constant of the system. We demonstrated that the single-photon absorption picture sustains up to hundreds of photons stored in the resonator. These findings also predict that for the high-impedance resonators with $Z_0\sim1/G_0$~\cite{stockklauser2017,childress2004}, the fine-structure constant approaches unity, and $n_r^c\sim 1$, so that the system switches from the single-photon 'particle' picture to the voltage-amplitude dominated regime immediately when the first photon is loaded into the cavity. Our results demonstrate a microwave version of the famous photoelectric effect and set an important milestone in bringing this key quantum optics concept to the experimental realm in the microwave domain. The present work opens up avenues of experiments using electron energy gain/loss spectroscopy in the microwave regime and shows the possibility of using a cavity-coupled DQD as the microwave energy harvester.

\begin{acknowledgments}
\par We acknowledge fruitful discussions with Adam Burke, Sven Dorsch, Martin Leijnse, Andreas Wacker and Peter Samuelsson and the financial support from NanoLund, Swedish Research Council (Dnr 2019-04111), the Foundational Questions Institute, a donor advised fund of Silicon Valley Community Foundation (FQXi-IAF19-07) and the Knut and Alice Wallenberg Foundation through the Wallenberg Center for Quantum Technology (WACQT).
\end{acknowledgments}
\bibliography{master}
\end{document}